\documentclass[prd,superscriptaddress,twocolumn,superscriptaddress,amsmath]{revtex4}
\usepackage{graphicx}
\usepackage{float}
\usepackage{epstopdf,cancel}
\usepackage{epsf,latexsym,bbm,euscript}
\usepackage{amssymb,amsmath,todonotes}
\usepackage{mathtools} %used in this file for the command \coloneqq -> :=
\usepackage{times,graphics}
\usepackage{xcolor}
\usepackage{soul}
\usepackage{epsfig,ulem}
\def\6{\langle}
\def\9{\rangle}

\newcommand\co{{\cal{O}}}

\newcommand\hn{{\hat{n}}}

\newcommand\sg{\textsl{g}}

\newcommand\bk{{\mathbf{k}}}

\newcommand\vv{{\vec{v}}}
\newcommand\vx{{\vec{x}}}

\newcommand\rtx{{{\rm tx}}}
\newcommand\rdet{{{\rm det}}}

\newcommand\cA{{\mathcal{A}}}

\def\half{\tfrac{1}{2}}

\newcommand{\pad}{\partial}
\newcommand{\be}{\begin{equation}}
\newcommand{\ee}{\end{equation}}
\newcommand{\ba}{\begin{eqnarray}}
\newcommand{\ea}{\end{eqnarray}}

\newcommand{\defeq}{\vcentcolon=}

\usepackage{enumitem}

 \usepackage{url,hyperref}
\hypersetup{colorlinks,linkcolor={blue!55!black},citecolor={red!50!black},urlcolor={blue!45!black}}

\begin{document}

\title{Large-scale optical interferometry in general spacetimes}
\author{Daniel~R.~Terno}
\affiliation{Department of Physics and Astronomy, Macquarie University, Sydney NSW 2109, Australia}
\affiliation{Shenzhen Institute for Quantum Science and Engineering, Southern University of Science and Technology, Shenzhen 518055, P. R. China}
\author{Giuseppe~Vallone}
\affiliation{Dipartimento di Ingegneria dell'Informazione, Universit\`{a} degli Studi di Padova, Padova 35131, Italy}
\affiliation{Istituto Nazionale di Fisica Nucleare (INFN) --- Sezione di Padova, Italy}
\affiliation{Dipartimento di Fisica e Astronomia, Universit\`a di Padova, via Marzolo 8, 35131 Padova, Italy}
\author{Francesco~Vedovato}
\affiliation{Dipartimento di Ingegneria dell'Informazione, Universit\`{a} degli Studi di Padova, Padova 35131, Italy}
\affiliation{Istituto Nazionale di Fisica Nucleare (INFN) --- Sezione di Padova, Italy}
\author{Paolo~Villoresi}
\affiliation{Dipartimento di Ingegneria dell'Informazione, Universit\`{a} degli Studi di Padova, Padova 35131, Italy}
\affiliation{Istituto Nazionale di Fisica Nucleare (INFN) --- Sezione di Padova, Italy}

\begin{abstract}
We introduce a convenient formalism to evaluate the  phase of  a light signal propagating on a general curved background. It allows to obtain a transparent  relation between the frequency-shift and the phase difference in large-scale optical interferometry for a general relativistic setting, as well as to derive compact expressions generalizing the Doppler effect in one-way and two-way schemes. Our recipe is easily applicable to stationary spacetimes, and in particular to the near-Earth experiments where  {the} geometry is described in the parametrized post-Newtonian approximation. As an example, we use it to evaluate the phase difference arising in the optical version of the Colella-Overhauser-Werner experiment.
\end{abstract}
\maketitle

\section{Introduction}
  Optical interferometry has been  crucial in   formulation and testing the special and general theories of relativity from the early times until present day~\cite{mm:887,gw-1,opt-bw,mtw,will:lrr}. Relative motion between the emitter and the detector, as well as propagation in a gravitational field, lead to changes in frequency --- the Doppler effect and  the gravitational red-shift, respectively --- that are  used in precision tests of relativity~\cite{mtw,will:lrr,is:938,pr:60,gpa:80,app:18, Delva2018, Herrmann2018}. The kinematic and gravitational effects are often competing, and strenuous efforts are made to separate the effects of relative motion from pure gravitational effects in both frequency and phase measurements~\cite{app:18,gpa:80,gpa,lisa-over,tdi}.

As long as the effects of quantum electrodynamics and/or quantum gravity are not important~\cite{qgo:03}, the propagation of the {electromagnetic} field is governed by the appropriate classical wave
equations on a fixed curved background~\cite{mtw,ueno75prl,harte:19}. For the minimally coupled electromagnetic field the vector potential $A^\mu$ satisfies the linear equation
 \be
 \Box A^\mu- R^\mu_{\ \nu} A^\nu=0 \ ,   \label{wave1}
 \ee
where $\Box\defeq\nabla^\mu\nabla_\mu$, with $\nabla_\mu$ the covariant derivative and $R^\mu_{\ \nu}$ the Ricci tensor that are associated with the background metric $\sg_{\mu\nu}$.

 The standard approach in classical and quantum optics \cite{opt-bw,qop} is to describe the wave propagation by means of geometric optics and, if necessary, its corrections.  These are derived
  by considering a decomposition of the vector potential as
  \begin{align}
  A^{\mu}(x) =  a^\mu e^{i\Phi(x)}+ e^{i\Phi(x)}\sum_{n\geqslant 1}^{{\infty}} {\tilde{\omega}}^{-n}\cA^\mu_n(x) \ , \label{asy_exp}
  \end{align}
   where {$\Phi$ is the phase (or eikonal function),}  the amplitudes $\cA_n$ are slowly-varying on the appropriate scales and the large parameter ${\tilde{\omega}}$ is  related to the peak frequency  of the solution~\cite{opt-bw,mtw,harte:19}.  The eikonal and the amplitudes can be determined from the equations for the coefficients of {the} various ${\tilde{\omega}}^{-n}$ terms  that are obtained by
 inserting this vector potential into the wave equation and imposing the Lorentz gauge $\nabla_\mu A^\mu = 0$.

In this work our motivation is twofold.  While the theory behind the calculation of the phase difference in optical interferometric experiments is long-established, not all of the implicit relations that are valid in a non-relativistic setting can be directly used in implementations involving arbitrary motion in a curved spacetime. This is so even  in the post-Newtonian and post-Minkowskian regimes~\cite{will:93,wp:book,bggst:15}, where  expressions for the time transfer function and the Doppler shift can be derived ~\cite{gpa-proposal,Blanchet2001,LaFitte2004,Hees2014}.
First, in Sec.~\ref{phase} by using the invariance of  {the} phase under arbitrary coordinate transformations, we arrive to an explicit expression for the phase-difference that depends on the proper emission time and frequency and that is valid in a general spacetime --- see Eq.~\eqref{cost_phase} and Eq.~\eqref{eq_phase_diff_main}. These equations are a straightforward application of known results, but their explicit expressions were never applied for optical interferometry in curved spacetimes.
Secondly, when applied to a particular large-scale optical interferometry set-up, namely the optical version of the Colella-Overhauser-Werner (COW) experiment~\cite{cole75prl}, as we present in Sec.~\ref{red-shift}, our recipe allows to simply relate the phase-difference --- see Eq.~\eqref{deltaPhi} --- to the well known frequency-shif, for which extended results are presented in literature~\cite{gpa-proposal,Blanchet2001, Hees2014}.

To make our presentation self-contained we show in Appendix~\ref{doppler} how the  frequency-shift, including the effects of both gravity and the relative motion (e.g., Doppler), can be easily obtained. This is essentially a re-statement of known results~\cite{gpa-proposal,Blanchet2001,LaFitte2004, Hees2014,Rakhmanov2006}, but it is presented in a form that is particularly convenient for the post-Newtonian analysis.

Both of these aspects are critical in  satellite-based gravitational experiments, either in the LISA gravitational wave antenna~\cite{lisa-over} or in the proposed optical tests of the Einstein Equivalence Principle (EEP)~\cite{zych11nco,sat:12}. In particular, this work provides the conceptual basis for the interferometric red-shift experiment aimed at testing the EEP proposed in Ref.~\cite{us-paper}, where a Doppler-cancellation scheme isolating the desired gravitational effect is introduced.

{\it Notation.---}We label the coordinates of a spacetime point as $x^\mu=(t,\vx)$, where $\vx$ designates a triple of spacelike coordinates $(x^1,x^2,x^3)$. These are given in a ``global'' coordinate frame, in contrast to any other ``local'' reference frame (fr) --- such as the one of emitter, mirror, detector, etc. --- that can be established, with the trajectories parametrized either by their proper time $\tau^\mathrm{fr}$ or by the global coordinate time $t$. We use the signature  $(-,+,+,+)$ for the metric. The quantity $r\defeq |\vx|= {\sqrt{(x^1)^2 + (x^2)^2 + (x^3)^2}}$ always stands for  length of the {Euclidean} vector $\vx$. Both co- and contravariant vectors on a three-dimensional curved space are designated by the boldface font, e.g., $\mathbf{k}$. Unless stated otherwise we use $G=c=1$.

\section{Phase evaluation
in a general relativistic setting} \label{phase}
{\it The laws of geometric optics.---}Within the domain of validity of geometric optics~\cite{opt-bw,harte:19}  the wave vector $k_\mu\defeq - \nabla_\mu \Phi \equiv  -\pad_\mu\Phi$ defines the propagation and the spatial periodicity of the wave. It is null (in all orders of the asymptotic expansion of Eq.~\eqref{asy_exp}) and thus it satisfies the eikonal equation:
\be
k^\mu k_\mu=\pad_\mu\Phi\pad^\mu\Phi=0 \ , \label{eq_eik}
\ee
which is a restatement of the null condition in terms of the phase function. Taking the gradient of the null condition,
$\nabla_\mu (k\cdot k)=0$, results in the
propagation equation for  the wave vector, that is
\be
k^\mu\nabla_\mu k^\nu=0 \ ,
\ee
where the geodesic is affinely parameterised.

 The three-dimensional hypersurfaces of constant  {$\Phi$} are null. In the high-frequency limit ${\tilde{\omega}}\to\infty$, these are the hypersurfaces of constant phase. The integral curves of $k^\mu$ form a twist-free null geodesic congruence. These geodesics {are the} light rays of geometric optics. {These rays can be also} be thought of as trajectories of fictitious photons that  {generate the} hypersurface $\Pi_\Phi$ of constant phase $\Phi$ and  at the same time are orthogonal to it {due to Eq.~\eqref{eq_eik}~\cite{mtw,ueno75prl}}. The eikonal equation [Eq.~\eqref{eq_eik}] is the Hamilton-Jacobi equation for massless particles on a given background spacetime. Its specific solution is determined by prescribing the phase on some initial spacelike hypersurface, e.g., $\Phi(t=t_0,\vx)$. In the caustic-free domain the value of $\Phi$ at some point $(t,\vx)$ is obtained by tracing the geodesic that passes through it to the point on the initial hypersuraface.

If we consider the vectorial nature of electromagnetic waves,
  the polarization vector is defined as $f^\mu\defeq a^\mu/\sqrt{a^\mu a^*_\mu}$. It is transversal to the null geodesic generated by $k^\mu$, and the Lorentz gauge condition implies the
parallel propagation equation for the polarization:
$f^\mu k_\mu=0$ and $k^\mu \nabla_\mu f^\mu=0$.

Solutions of the Hamilton-Jacobi equation are particularly simple in   stationary spacetimes, where
the metric tensor is independent of the time coordinate. In such spacetimes existence of the timelike Killing vector $\pad_0$, such that $\pad_0\sg_{\mu \nu}=~0$, ensures  that $\omega_0=-\big(k[t_0,\vx_0]\big)_0$ is constant along the geodesic.

If the frequency is fixed in the proper frame of a static observer then $\omega_0$ has same value on all geodesics that emanate
 from $\vx_0$. Given the orthonormal tetrad $e^{(A)}_\mu$,  $A=0,\ldots 3$, the conserved frequency is  $\omega_0=-e^{(A)}_0k_A=\mathrm{const}$, where $k_A$ are components of the wave 4-vector in the proper frame of the observer. The 4-velocity of the observer defines its time axis. For a static observer in a stationary spacetime  it is $e^{(0)\mu}=\delta^\mu_0/\sqrt{-\sg_{00}}$, and thus by orthogonality of the tetrad $e^{(I)}_0=0$, for $I=1,2,3$. As a result the conserved frequency depends only on the proper frequency and the metric.

Static observers  in stationary spacetimes follow the congruence of timelike Killing vectors $\pad_0$ that defines a projection from the space-time manifold $\cal{M}$ onto a three-dimensional space $\Sigma_3$. Using the Landau-Lifshitz formalism~\cite{ll:2,fl:82,bt:11} the spacetime domain is foliated by the hypersurfaces of simultaneity $\Sigma_3(t)$ with respect to the static observer. This time $t$ is the universal time that we use below. The foliation introduces the time-independent spatial metric $\gamma_{mn}$ that determines   geometric properties of the three-dimensional space $\Sigma_3$. The three coordinates of the point $x^\mu=(x^0=t,\vx)\in\Sigma_3(t)$ are just the triple $\vx=(x^1,x^2,x^3)$. Both spatial vectors and covectors on this space are obtained by simply retaining the spatial components, as $\bk^m=k^m$, and $\bk_m=\gamma_{mn}\bk^n\equiv k_m$.

Intersection of the hypersurface of simultaneity $\Sigma_3(t)$ with the hypersurface of the constant phase $\Pi_\Phi$ results in the the instantenious two-dimensional surface $\Pi_2(t;\Phi)$ of the constant phase of the wave front.  While the four-vector $k$ is tangent to the world line of a fictitious photon and orthogonal to the null hypersurface $\Pi_\Phi$, its three-dimensional projection  $\bk$ is perpendicular to
 the surface of constant phase $\Pi_2$ and tangent to the light ray in the space $\Sigma_3$. In static spacetimes the geodesic equation and the evolution equation of polarization can be conveniently written in a three-dimensional form~\cite{fl:82, bt:11}. If $(t,\vx)$ is connected to $(t_0,\vx_0)$ by a null geodesics, then the coordinate travel time depends only on the spatial coordinates via some function $T(\vx;\vx_0)$ such that $t-t_0=T(\vx;\vx_0)$.

Thus the leading term of the asymptotic expansion  leads to the  {\it three laws of geometric optics} \cite{opt-bw,mtw,harte:19}. Namely:
\begin{enumerate}[label=(\roman*)]
\item Fields propagate along   null geodesics. In stationary spacetimes their propagation can be visualized as
{the} advance  of a two-dimensional surface of constant phase in  three-dimensional space that is  guided by the light rays.

\item Polarization is parallel-transported along the rays.

\item Intensity satisfies the inverse-area conservation law.
\end{enumerate}
 The superposition of two or more light waves results in the appearance of interference. Phases, intensities and polarizations of the individual waves are calculated in the approximation of
 geometric optics according to the above rules  (i)-(iii).

\begin{figure}[!t]
    \centering
   \includegraphics[width=0.37\textwidth]{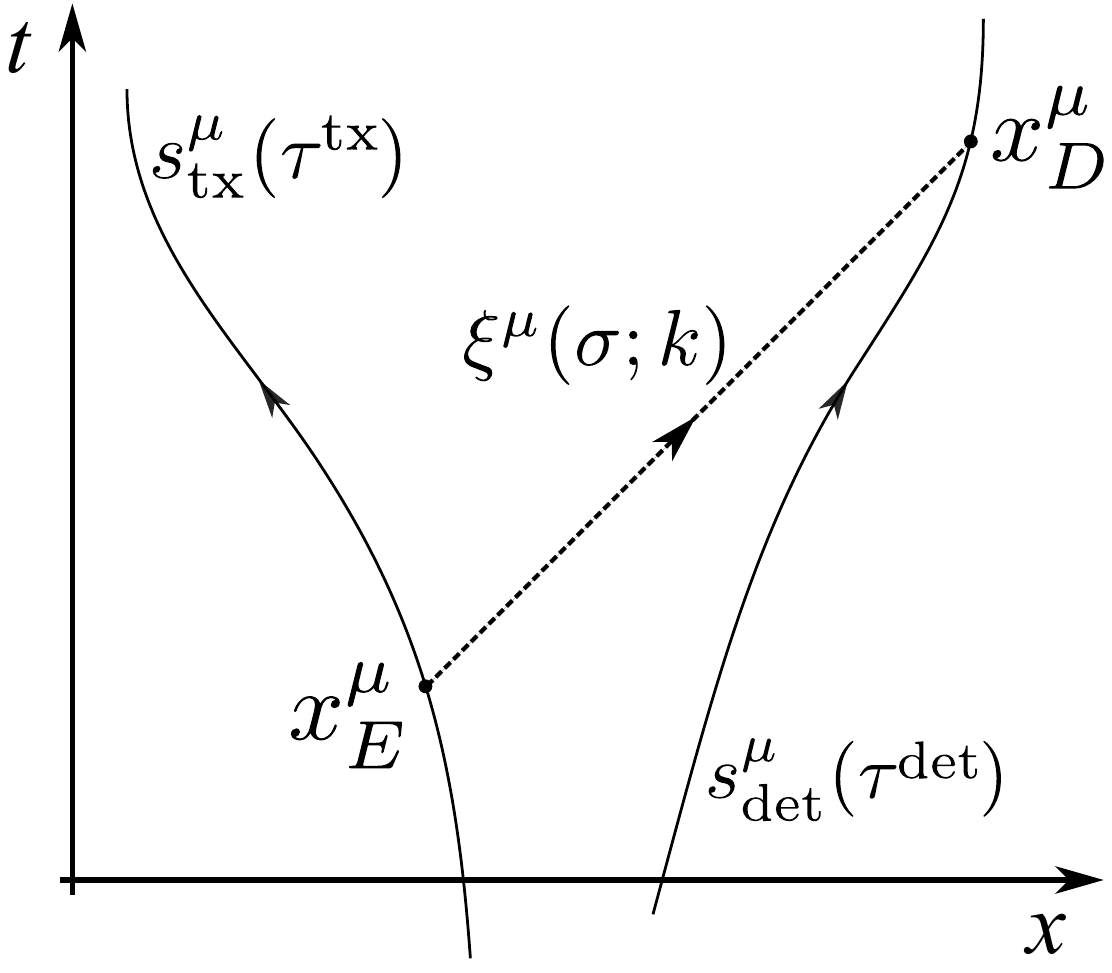}
    \caption{A schematic representation of the world lines of the emitter (trajectory $s^\mu_\rtx$), the detector (trajectory $s^\mu_\rdet$) and the light ray
   {$\xi^\mu$}  between the emission event $E$ and the detection event $D$ (dotted straight line).}
    \label{fig:trajectories}
\end{figure}

 \medskip
{\it Phase evaluation.---}Consider now a single null geodesic   segment that connects two points --- $(t_E, \vx_E)$ and $(t_D, \vx_D)$ --- belonging to two timelike trajectories which are  parametrized  by their respective proper times, as sketched in Fig.~\ref{fig:trajectories}. One, $s^\mu_{\rm det}(\tau^{\rm det})$, represents a detector and the other, $s^\mu_{\rm tx}(\tau^{\rm tx})$,   represents the transmitter.  In the frame of the point-like transmitter, the phase of the emitted signal at the emission instant $\tau_E^{\rm tx}$ is $\Phi_E(\tau^{\rm tx}_{{E}})$ and can be written as
\be
\Phi_E(\tau^{\rm tx}_{ {E}})=-\omega_{\rm tx}(\tau^{\rm tx}_{ {E}}-\tau^{\rm tx}_0)\ \label{phiE},
\ee
 where we assume a constant proper frequency $\omega_{\rm tx}$ and  $\tau^{\rm tx}_0$ determines the initial phase.
 We recall that the local frequency
$\omega_\mathrm{fr}$ of the optical signal with wave vector $k$ in some frame that is moving with the four-velocity $u^\mu_\mathrm{fr}(\tau^\mathrm{fr})$  is $\omega_\mathrm{fr}=-u^\mu_\mathrm{fr} k_\mu$.

Since the optical phase is constant along a spacetime trajectory of the photon, the phase of the signal that is detected at the spacetime location  $x^\mu_D=s^\mu_{\rm det}(\tau_D^{\rm det})$  equals to the phase of the emitted signal at $x^\mu_{E}$.
Given the detection at $x^\mu_D$ of the signal with $k^\mu$, the coordinates of the emission event $x^\mu_E = s^\mu_{\rm tx}(\tau_E^{\rm tx})$ are found as the intersection of the
 backward propagated geodesic from the detection point with the wordline of the emitter, that is
\be
s^\mu_{\rm tx}(\tau_E^{\rm tx})=\xi^\mu(\sigma; k) \ ,
\ee
where $\sigma$ is the affine parameter
 and $\xi^\mu(\sigma; k)$ is the spacetime trajectory of the photon. Using (i) the detected phase in terms of the properties of the emitter  is given by
\be
\Phi_D(\tau_D^{\rm det})=\Phi_E\left(\tau_E^{\rm tx}(k)\right) \ . \label{cost_phase}
\ee

 When it does not lead to confusion we simply write $\tau_E \equiv \tau_E^{\rm tx}$ and  $\tau_D \equiv \tau_D^{\rm det}$ implying that we use the explicit form of the trajectories $s^\mu_\rtx(\tau^\rtx)$ and   $s^\mu_\rdet(\tau^\rdet)$ to establish the relationship between the proper times of the emission and detection of a signal in the respective frames. For a trajectory that consists of several geodesic segments the time procedure remains the same, but, in addition to the flight time, the phase changes at the nodes (such as $\pi$ phases at the reflections) should be added to the final expression for the   phase.

 We notice that, from the constancy of phase [Eq.~\eqref{cost_phase}], it is possible to obtain the  well-known  relationship~\cite{Blanchet2001,LaFitte2004,Hees2014} between the detected frequency $\omega_\rdet$ and the transmitted one $\omega_\rtx$, according to
\be
\omega_\rdet=-{\frac{d\Phi_D}{d\tau_D}}= -\frac{d\Phi_E}{d\tau_D}=\left(- \frac{d\Phi_E}{d\tau_E} \right)\frac{d\tau_E}{d\tau_D}=\omega_\rtx \frac{d\tau_E}{d\tau_D} \ ,  \label{freq0}
\ee
where $d\tau_E/d\tau_D$ can be obtained by differentiating the coordinate-time transfer~\cite{LaFitte2004}
\begin{equation}
    t_D(\tau_D) - t_E(\tau_E) = \mathbb{T}\big(s_\rdet(\tau_D);s_\rtx(\tau_E)\big)  \ .  \label{tedgen}
\end{equation}
In the above equation, $\mathbb{T}$  is the function that implicitly  captures the relation between the proper time of emission and detection in the respective frames (see Appendix~\ref{doppler} for more details and the explicit expressions in the case of  post-Newtonian  expansion).
In a stationary spacetime the $\mathbb{T}$ function depends only on spatial coordinates: in this case, $\mathbb{T}=T(\vx_D;\vx_E)$ represents the interval of coordinate time that takes  a null particle to travel from $\vx_E$ to $\vx_D$.

Consider now a two-beam interference where the beams {arrive} to the detector via two different paths. Here we assume that the acquired phase is only due to space-time propagation and not to additional phase shifts due to optical elements (such as mirrors).
Let the first and the second beams that arrive at $x_D$ to have the  wave vectors $k_1$ and  $k_2$, respectively. Then, the  phase difference at $x_D$ is given by

\be
\Delta\Phi(\tau_D)=\Phi_E\big(\tau_E( {k_2})\big)-\Phi_E\big(\tau_E( {k_1})\big) \ , \label{eq_phase_diff_main}
\ee
where, in general,
$\tau_E(k_1) \neq \tau_E(k_2)$
due to the  back-propagation, which is different for the two paths. The above relation shows that the phase difference measured at the space-time location $x^\mu_D$ is equal to the difference between the phases at the transmitter evaluated at the two emission times $\tau_E(k_1)$ and $\tau_E(k_2)$.

\section{Application to large-scale optical interferometry}  \label{red-shift}

Measurements of the gravitational red-shift   provide one of the fundamental tests of general relativity and metric theories of gravity in general~\cite{will:lrr,gpa:80,mtw,will:93}. Possible violations of the equivalence principle (and, specifically of the assertion that outcomes ``of any local non-gravitational experiment is independent of where and when in the universe it is performed''~\cite{will:lrr}) can occur as a result of a subtle interplay between different sectors of the Standard Model and its extension. In this regard a purely optical test based on large-scale optical interferometry provides a new type of a probe.

\begin{figure}[t]
    \centering
    \includegraphics[width=0.4 \textwidth]{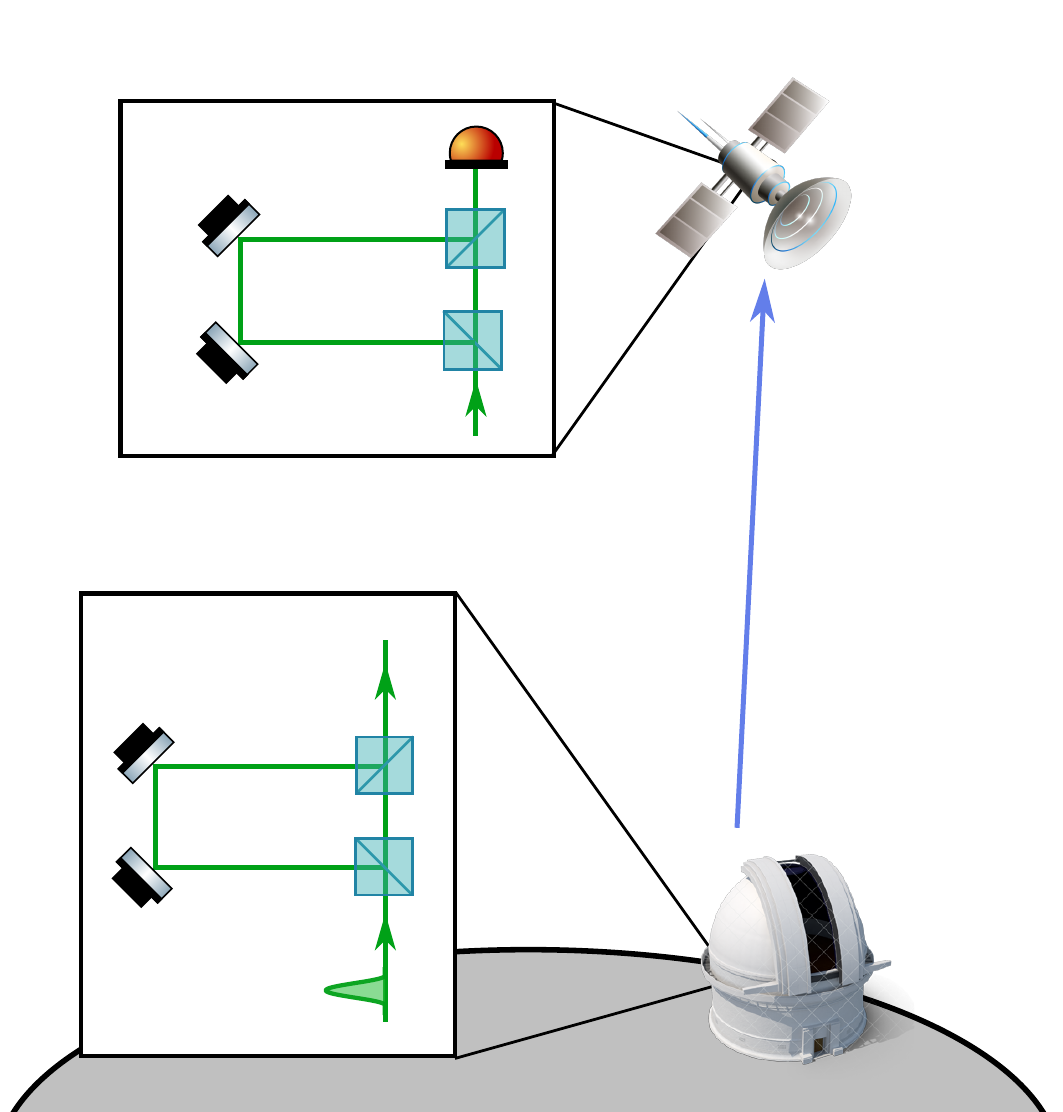}
    \caption{Schematic representation of the interferometric red-shift measurement.
    For the stationary emitter and detector the only relevant effect is the difference in the coordinate time intervals that correspond to the same proper delay time $\tau_l$.}
    \label{fig:setup}
\end{figure}

A basic version of the interferometric red-shift experiment that tests the ``where" part of the above assertion and is known as the ``optical-COW''~\cite{zych11nco,sat:12}  is represented in Fig.~\ref{fig:setup}. A light pulse is coherently split into two on the ground by using an interferometer of temporal imbalance $\tau_l = l/c$, with $l$ the length of the optical delay line.
The two pulses are recombined at the satellite by using another interferometer with the same imbalance $\tau_l$.  Since the  two interferometers sit at different gravitational potentials there will be a gravitational phase difference between the two interfering paths. It is estimated as~\cite{sat:12,bggst:15}
\begin{equation}
\varphi_{\rm gr} = \frac{2\pi}{\lambda}\frac{gh l }{c^2} \ ,
\label{phigr}
\end{equation}
where  $g$ is the Earth's gravity, $h$ the satellite altitude and $\lambda = 2\pi c/\omega$ the sent wavelength.
Putting the emitter on a ground-station~(GS) and the detector on a low-Earth-orbit spacecraft~(SC), $\varphi_{\rm gr}$ results of the order of few radians, supposing, as in Ref.~\cite{sat:12}, a delay of $l = 6$~km, $\lambda = 800$~nm and $h = 400$~km. However, if the relative motion of the emitter and the detector is taken into account~\cite{bggst:15}, then the first-order Doppler effect is roughly $10^5$ times stronger than the desired signal $\varphi_{\rm gr}$.

Now we provide a careful evaluation of the phase difference and show its relation to the frequency-shift. This analysis underlines the Doppler cancellation scheme that is proposed in Ref.~\cite{us-paper} to optically bound the violation of the EEP down to $10^{-5}$, a precision of the same order of the best absolute results obtained so far~\cite{Delva2018,Herrmann2018}.

 \begin{figure}[b] \centering
    \includegraphics[width=0.25\textwidth]{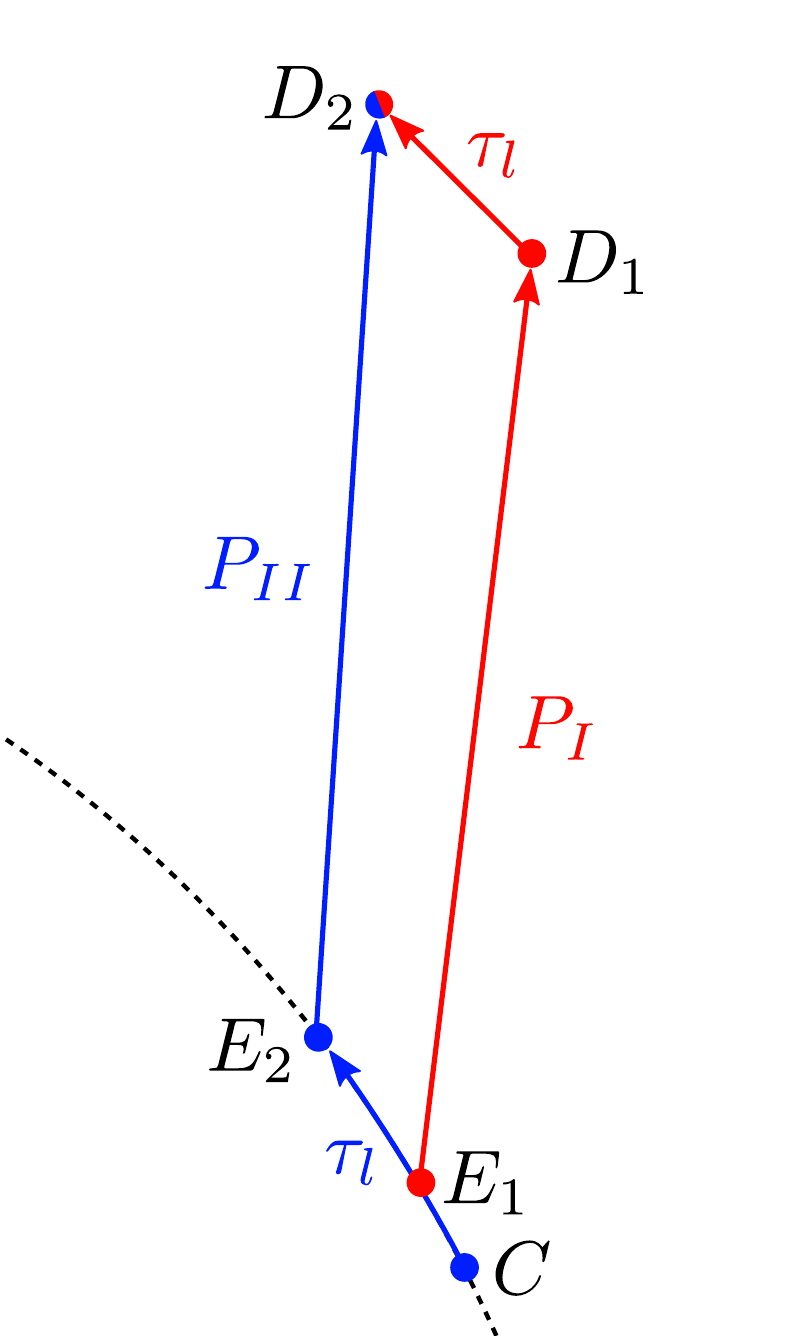}
    \caption{Sequence of events in the experiment.
    The red line $P_I$ represents the signal that is delayed on board of the SC, and the blue line $P_{II}$ is the one delayed at the GS prior to transmission. The proper time intervals between the events $D_1$ and $D_2$ (at the SC) and $C$ and $E_2$ (at the GS) are both equal to the proper temporal imbalance $\tau_l$.}
    \label{fig:space}
\end{figure}

The key events of the ``optical-COW'' experiment are depicted on Fig.~\ref{fig:space}. Due to the Earth motion, the two pulses that recombine at the SC at the point $D_2$ of the diagram leave the GS at two different times: the signal that takes the short path in the GS-interferometer leaves the ground at $\tau_{E_1}$, while the pulse that takes the long arm is delayed by $\tau_l$ and departs at $\tau_{E_2}$. The interference condition {for the paths $P_I$ and $P_{II}$} at $D_2$ implies that the phase of the delayed pulse has to be evaluated at the instant $\tau_C := \tau_{E_2} - \tau_l$.  Indeed, by taking into account the presence of the delay-lines (d.l.) and of the back-propagation (Eq.~\eqref{cost_phase}) in the given path, we have that

Due to the Earth motion, the two pulses that recombine at the SC at the point $D_2$ of the diagram leave the GS at two different times: the signal that takes the short path in the GS-interferometer leaves the ground at $\tau_{E_1}$, while the pulse that takes the long arm is delayed by $\tau_l$ and departs at $\tau_{E_2}$. The interference condition {for the paths $P_I$ and $P_{II}$} at $D_2$ implies that the phase of the delayed pulse has to be evaluated at the instant $\tau_C := \tau_{E_2} - \tau_l$.
Indeed, by taking into account the presence of the delay-lines (d.l.) and of the back-propagation (Eq.~\eqref{cost_phase}) in the given path, we have that
    \begin{align}
    \Phi_{D}(\tau_{D_2}|P_I) &\overset{\rm d.l.}{=} \Phi_{D}(\tau_{D_1} )
    \overset{\eqref{cost_phase}}{=} \Phi_{E}(\tau_{E_1})
    \\
    \Phi_{D}(\tau_{D_2}|P_{II}) &\overset{\eqref{cost_phase}}{=} \Phi_{E}(\tau_{E_2})
    \overset{\rm d.l.}{=} \Phi_{E}(\tau_{C} ) \ ,
\end{align}
where $\tau_{D_1} := \tau_{D_2} - \tau_l$
and $\tau_{C} := \tau_{E_2} - \tau_l$.

Hence, the phase difference detected at the SC, according to Eq.~\eqref{eq_phase_diff_main} and by using Eq.~\eqref{phiE}, is
\be
\Delta \Phi=\Phi_E(\tau_{E_1})-\Phi_E(\tau_{C})=-\omega_\rtx(\tau_{E_1}-\tau_{C}) \ .      \label{phase1}
\ee
The above relation has a clear physical meaning: the phase difference measured on the satellite is obtained by the difference of the phases at the transmitter evaluated at the (proper) times $\tau_{E_1}$ and $\tau_{C}$, which correspond to the emission times of the light detected at the event $D_2$ and that followed the path $P_I$ and $P_{II}$ respectively.
The above  difference of phases at the transmitter is simply obtained as the proper frequency $\omega_{\rm tx}$ multiplied by the time difference $\tau_{C}-\tau_{E_1}$.

We will now show that for a sufficiently short proper delay time $\tau_l$   the phase {difference} in this scheme is proportional to the frequency difference {$\omega_{\rm det} - \omega_{\rm tx}$}.  {In fact, by} defining $\Delta \tau_E:=\tau_{E_2}-\tau_{E_1}$, we have that
$\tau_{E_1} - \tau_C = \tau_l - \Delta \tau_E$.
The proper time difference $\Delta\tau_E$ can be found from the relation
 \be
   \Delta \tau_E\simeq\frac{d\tau_E}{d\tau_D}\tau_l
 \ee
if only the leading term in $\tau_l$ is kept.
 Hence,
 \begin{equation}
  {\Delta\Phi~{\simeq} - \omega_{\rm tx} \tau_l \left(1 - \frac{d\tau_E}{d\tau_D} \right) {\simeq} ~ \left(\omega_{\rm det} - \omega_{\rm tx}\right) \tau_l} \ , \label{deltaPhi}
 \end{equation}
 where the last equality follows from Eq.~\eqref{freq0}.  This expression is the first-order approximation in $\tau_l$ and  is valid if both $v_E\gg a_E (dt_E/d\tau_E)\tau_l$ and $v_D\gg a_D (dt_D/d\tau_D)\tau_l$ hold, where $a_E$ and  $a_D$ are accelerations of the GS and SC at the emission and the detection times, respectively. In the case of the post-Newtonian metric, the {explicit} relation between $\omega_{\rm det}$ and $\omega_{\rm tx}$ is given by the well known one-way frequency-shift~\cite{gpa-proposal} (see also Eq.~\eqref{freqgp} of Appendix~\ref{doppler}).

\section{Summary}

Many of the implicit assumptions of optical interferometry, such as the relationships between distance and time, or even the logical consistency of the assumption that two interfering beams have the same central frequency,  are not valid in the relativistic setting. However, the primary interpretation of the phase difference as arising from the difference in the emission times allows to obtain compact expressions that provide the conceptual basis for analysis of large-scale interferometric experiments in general spacetimes.

 The constancy of the phase on propagating wave surfaces in the geometric optics approximation allows for a simple generalization of the Doppler effect affecting the frequency to curved spacetimes. The resulting expression  does not require transformations between reference frames for its use.

 Moreover, in the limit of short delay times, the phase  difference in the described interferometric measurement of the gravitational red-shift  is proportional to this frequency difference, confirming the original order of magnitude estimates and the unavoidable dominance of the first-order Doppler effect in the one-way interferometric red-shift experiment.

\acknowledgments

The work of DRT is supported by the  grant   FA2386-17-1-4015 of AOARD. Useful discussions with Alex Ling and Alex Smith are gratefully acknowledged. {Fig.~3 was designed by Alex Smith.
 \appendix

 \section{Frequency-Shift Evaluation}
\label{doppler}

 \subsection{Emission-Detection frequency-shift}

{On} a generic background  we obtain the frequency-shift {of Eq.~\eqref{freq0}} by differentiating Eq.~\eqref{tedgen} and solving
 \be
\frac{dt_D}{d\tau_D}=\frac{dt_E}{d\tau_E}\frac{d\tau_E}{d\tau_D}+\frac{\pad\mathbb{T}}{\pad s^\mu_\rdet}u^\mu_D+    \frac{\pad\mathbb{T}}{\pad s^\mu_\rtx}u^\mu_E \frac{d\tau_E}{d\tau_D}
\ee
 for $d\tau_E/d\tau_D$.  Here  the four-velocities are
\begin{align}
&u^\mu_D=\frac{ds^\mu_\rdet}{d\tau_D} \ ,   \quad u^\mu_E =\frac{ds^\mu_\rtx}{d\tau_E}\ , \\ &\frac{dt_D}{d\tau_D}  = u^0_D\ ,  \quad \frac{dt_E}{d\tau_E}=u^0_E\ ,
\end{align}
and the proper time is related to the coordinate time via
\be
d\tau=\sqrt{|\sg_{00}|-\sg_{0k}v^k-\sg_{kl}v^kv^l} \ dt \ .
\ee
The above expression allows to evaluate the frequency-shift in a generic background by noting that Eq.~\eqref{freq0} can be re-written as
\be
\omega_\rdet=\omega_\rtx\frac{d\tau_E}{dt_E}\frac{dt_E}{dt_D}\frac{dt_D}{d\tau_D} \ .          \label{freq1}
\ee

 A more explicit expression is possible in a stationary spacetime, where
 Eq.~\eqref{tedgen} reduces to
\be
t_D=t_E+T(\vx_D;\vx_E) \ , \label{tD_stationary}
\ee
where $T(\vx_D;\vx_E)$ is the interval of coordinate time that takes  a null particle to travel from $\vx_E$ to $\vx_D$, yielding
\be
1=\frac{dt_E}{dt_D}+ \frac{\pad T}{\pad \vx_D}\cdot\vv_D  +\frac{\pad {T}}{\pad \vx_E}\cdot\vv_E  \frac{dt_E}{dt_D} \ .
\ee

For the near-Earth experiments the spacetime is well-approximated by the post-Newtonian expansion \cite{mtw,will:93,wp:book}.
The paramterized post-Newtonian formalism  admits a broad class of metric
theories of gravity, including general relativity as a special
case. The key small parameter is $\epsilon^2\sim GM/c^2r\sim v^2/c^2$, where
$v$ is the velocity of a massive test particle or of some
component of a gravitating body. The metric including the leading post-Newtonian terms (up to the second order in $\epsilon$) is stationary and it is given by
\be
\sg_{00}=-1+ 2 U \ , \qquad \sg_{ij}=\delta_{ij}\big(1+ 2 U\big) \ , \label{eq_ppn_metric}
\ee
with $U \equiv U(\vx)\defeq {GM}Q(r,\theta)/rc^2$  denoting the gravitational potential around the Earth, including the quadrupole term
\be
Q(r,\theta) :=1-\half J_2\frac{R^2}{r^2}(3\cos^2\theta-1)\ ,
\ee
where $J_2=1.083\times 10^{-3}$ is the normalized quadrupole moment and  the higher-order terms \cite{allen,gnss}.  $R$ is the Earth equatorial radius. Given  the established
bounds on the post-Newtonian parameter $\gamma$~\cite{will:lrr} we set $(1+\gamma) = 2$ in the metric.

For a spherically-symmetric Earth in the leading  post-Newtonian expansion, the photon time-of-flight is given by~\cite{will:93,wp:book}
\begin{equation}
     T^{\rm PPN}(\vx_D;\vx_E) = T_0(\vx_D;\vx_E)+ T_2(\vx_D;\vx_E) \ , \label{shapiro}
\end{equation}
where $T_0(\vx_j;\vx_i) := |\vx_j - \vx_i|$ is the flat spacetime result and the leading post-Newtonian term is
\be
    T_2(\vx_j; \vx_i) := 2M\ln\frac{r_j+\vx_j\cdot\hn_{ij}}{r_i+\vx_i\cdot\hn_{ij}} \ , \label{T_2}
\ee
with $r_i:= |\vx_i|$ the Euclidean length and
 \begin{equation}
 \hn_{ij} \defeq\frac{\vx_j-\vx_i}{|\vx_j-\vx_i|}=\frac{\vx_j-\vx_i}{r_{ij}}
 \end{equation}
is the Euclidean unit vector along the Newtonian propagation direction.
For what follows, we note that
\begin{equation}
    \frac{\displaystyle\pad T_0}{\displaystyle\pad \vx_j} (\vx_j;\vx_i) = \hn_{ij} = - \frac{\displaystyle\pad T_0}{\displaystyle\pad \vx_i} (\vx_j;\vx_i) \ ,
\end{equation}
{and} that $T^{\rm PPN}(\vx;\vx_0)$ correctly satisfies (up to the second order in $\epsilon$) the  eikonal equation in Eq.~\eqref{eq_eik}  with the  metric of Eq.~\eqref{eq_ppn_metric}, i.e.,
\begin{align}
    \left|\frac{\partial T_0}{\partial \vx}(\vx; \vx_0)\right|^2 &= 1 \ , \quad \frac{\partial T_0}{\partial \vx} (\vx; \vx_0) \cdot \frac{\partial T_2}{\partial \vx} (\vx; \vx_0) = 2 U(\vx) \  . \label{eq_eik_PPN}
\end{align}

Then, in the leading post-Newtonian approximation using Eq.~\eqref{shapiro} we obtain
\be
\frac{dt_E}{dt_D}=\frac{1-\hn_{ED}\cdot\vv_D -\frac{\displaystyle\pad T_2}{\displaystyle\pad \vx_D}\cdot\vv_D}{1- \hn_{ED}\cdot\vv_E+
\frac{\displaystyle\pad T_2}{\displaystyle\pad \vx_E}\cdot\vv_E}=
\frac{1-\hn_{ED}\cdot\vv_D }{1- \hn_{ED}\cdot\vv_E}+\co(\epsilon^3) \ . \label{eq_ratio_diff}
\ee
and the   metric gives
\be
d\tau=\sqrt{1-2U-v^2}dt+\co(\epsilon^4) \ ,
\ee
allowing to write, for example,
\begin{equation}
    d\tau_E=\sqrt{1-2U_E-v_E^2} \ dt_E
\end{equation}
   so that at the second order in $\epsilon$ we obtain the standard expression for the one-way (1w) frequency-shift~\cite{gpa,gpa:80,gnss, gpa-proposal}
\begin{align}
    \left.\frac{\omega_\rdet}{\omega_\rtx}\right|_{\rm 1w}&=\sqrt{\frac{1-2U_E-v^2_E}{1-2U_D-v_D^2}}\left(\frac{1-\hn_{ED}\cdot\vv_D }{1- \hn_{ED}\cdot\vv_E}\right)+\co(\epsilon^3) \nonumber\\
    &= 1 + \hn_{ED}\cdot (\vv_E - \vv_D) \nonumber\\
    &\quad+ U_D - U_E + \frac{1}{2} (v_D^2 - v_E^2) + (\hn_{ED}\cdot \vv_E)^2 \nonumber\\
    \label{freqgp}
    &\quad\quad - (\hn_{ED}\cdot\vv_D)(\hn_{ED}\cdot\vv_E) + \co(\epsilon^3) \ .
\end{align}
The first-order contribution provides the usual Doppler effect depending on the relative velocity between the transmitter and the detector: \begin{equation}\left.\frac{\omega_{\rm det}}{\omega_{\rm tx}}\right|_{\rm 1w}^{(\epsilon)} = \hn_{ED} \cdot (\vv_E - \vv_D) \ .
\end{equation}

 \subsection{Emission-Reflection-Detection frequency-shift}
In many practical situations, prior to the detection, the beam is reflected  by a moving mirror. This is for example the case of the experiment realized in  Ref.~\cite{padova:16} and also the situation analyzed in  Ref.~\cite{Rakhmanov2006}. We illustrate the analysis in this setting by considering the motion in a stationary spacetime.
  The path from the emission event $E$ to the detection $D$  now comprises two geodesic segments, $E \rightarrow M \rightarrow D$, where $M$ stands for mirror. Hence, Eq.~\eqref{tD_stationary} is replaced by a pair of equations
\begin{align}
    t_D &= t_M + T(\vx_D; \vx_M) \ , \label{eq_td_mirror} \\
    t_M &= t_E + T(\vx_M; \vx_E) \ , \label{eq_tm_mirror}
\end{align}
that conveniently decompose the flight time $t_D-t_E$.
Now, the constancy of the phase allows to write $\omega_\rdet/\omega_\rtx = d\tau_E/d\tau_D$ --- see Eq.~\eqref{freq0} ---
and the analogue to Eq.~\eqref{freq1} now passes through $M$ as
\begin{equation}
    \omega_\rdet = \omega_\rtx \frac{d\tau_E}{d t_E} \frac{d t_E}{d t_M} \frac{d t_M}{d t_D} \frac{d t_D}{d \tau_D} \ .
\end{equation}
In the post-Newtonian approximation the two central ratios in the equation above are evaluated analogously to Eq.~\eqref{eq_ratio_diff} by using Eqs.~\eqref{eq_td_mirror}-\eqref{eq_tm_mirror}, yielding
\begin{align}
    \frac{d t_E}{d t_M} &=
\frac{1-\hn_{EM}\cdot\vv_M }{1- \hn_{EM}\cdot\vv_E}+\co(\epsilon^3) \\
    \frac{d t_M}{d t_D} &=
\frac{1-\hn_{MD}\cdot\vv_D }{1- \hn_{MD}\cdot\vv_M}+\co(\epsilon^3)
\end{align}
and thus the two-way (2w) frequency-shift~\cite{gpa-proposal} is (up to $\co(\epsilon^3)$)
\begin{widetext}
\begin{align}
    \left.\frac{\omega_\rdet}{\omega_\rtx}\right|_{\rm 2w}&=\sqrt{\frac{1-2U_E-v^2_E}{1-2U_D-v_D^2}} \  \left(\frac{1-\hn_{EM}\cdot\vv_M }{1- \hn_{EM}\cdot\vv_E} \right) \left( \frac{1-\hn_{MD}\cdot\vv_D }{1- \hn_{MD}\cdot\vv_M} \right) + \co(\epsilon^3) \nonumber\\
    &= 1 + \hn_{EM}(\vv_E - \vv_M) + \hn_{MD} (\vv_M - \vv_D) \ + U_D - E_E + \frac{1}{2} (v_D^2 - v_E^2) \nonumber\\
    &\quad - (\hn_{EM}\cdot\vv_M)(\hn_{EM}\cdot\vv_E) + (\hn_{EM}\cdot\vv_E)^2  - (\hn_{MD}\cdot\vv_D)(\hn_{MD}\cdot\vv_M) + (\hn_{MD}\cdot\vv_M)^2 \nonumber \\
    &\quad\quad+ \left[\hn_{EM}\cdot(\vv_M - \vv_E)\right]\left[\hn_{MD}\cdot(\vv_D - \vv_M)\right] + \co(\epsilon^3) \ .
\end{align}
\end{widetext}
If the emission and detection occur at the same ground station, then the first factor on the  right-hand-side of above equation equals to unity.


\begin{thebibliography}{99}

\bibitem{mm:887} A. A. Michelson and E. W. Morley, \textit{On the relative motion of the Earth and the luminiferous ether}, \href{http://www.ajsonline.org/content/s3-34/203/333}{Am. J. Science \textbf{34}  (203), 333 (1887)}.

\bibitem{gw-1} LIGO Scientific Collaboration and Virgo Collaboration, \textit{Observation of gravitational waves from a binary black hole merger}, \href{https://doi.org/10.1103/PhysRevLett.116.061102}{Phys. Rev. Lett. \textbf{116}, 061102 (2016)}.

\bibitem{opt-bw} M. Born and E. Wolf, \textit{Principles of Optics}, 7th ed., (Cambridge University Press, Cambrdge, England, 1999).
\bibitem{mtw} C. W. Misner, K. S. Thorn, and J. A. Wheeler, \textit{Gravitation},
(Freeman, San Francisco, 1973).

\bibitem{will:lrr}C.~M.~Will, \emph{The Confrontation between General Relativity and Experiment}, \href{https://doi.org/10.12942/lrr-2014-4}{Living Rev. Relativity \textbf{17}, 4 (2014)}.

\bibitem{is:938} H. E. Ives and G.R. Stilwell, \textit{An experimental study of the
rate of a moving atomic clock}, \href{https://doi.org/10.1364/JOSA.28.000215}{J. Opt. Soc. Am. \textbf{28}, 215 (1938)}.

\bibitem{pr:60} R.~V.~Pound and G.~A.~Rebka, Jr., \emph{Apparent Weight of Photons}, \href{https://doi.org/10.1103/PhysRevLett.4.337}{Phys. Rev. Lett. \textbf{4}, 337 (1960)}.

\bibitem{app:18}
N.~Ashby, T.~E.~Paker, and B.~R.~Patla,
\emph{A null test of general relativity based on a long-term comparison of atomic transition frequencies},
\href{https://doi.org/10.1038/s41567-018-0156-2}{Nature Phys. {\bf 14},  822-826 (2018)}.

\bibitem{Delva2018}
P.~Delva, N.~Puchades, E.~Sch\"{o}nemann, F.~Dilssner, C.~Courde, S.~Bertone, F.~Gonzalez, A.~Hees, Ch.~Le~Poncin-Lafitte, F.~Meynadier, R.~Prieto-Cerdeira, B.~Sohet, J.~Ventura-Traveset, and P.~Wolf,
\emph{Gravitational Redshift Test Using Eccentric Galileo Satellites},
\href{https://doi.org/10.1103/PhysRevLett.121.231101}{Phys. Rev. Lett. {\bf 121}, 231101 (2018)}.

\bibitem{Herrmann2018}
S.~Herrmann, F.~Finke, M.~L\"{u}lf, O.~Kichakova, D.~Puetzfeld, D.~Knickmann, M.~List, B.~Rievers, G.~Giorgi, C.~G\"{u}nther, H.~Dittus, R.~Prieto-Cerdeira, F.~Dilssner, F.~Gonzalez, E.~Sch\"{o}nemann, J.~Ventura-Traveset, and C.~L\"{a}mmerzahl,
\emph{Test of the Gravitational Redshift with Galileo Satellites in an Eccentric Orbit},
\href{https://doi.org/10.1103/PhysRevLett.121.231102}{Phys. Rev. Lett. {\bf 121}, 231102 (2018)}.

\bibitem{gpa:80}
R.~F.~C.~Vessot, M.~W.~Levine, E.~M.~Mattison, E.~L.~Blomberg, T.~E.~Hoffman, G.~U.~Nystrom, B.~F.~Farrel, R.~Decher, P.~B.~Eby, C.~R.~Baugher, J.~W.~Watts, D.~L.~Teuber, and F.~D.~Wills,
\textit{Test of relativistic gravitation with a Space-borne hydrogen maser}, \href{https://doi.org/10.1103/PhysRevLett.45.2081}{Phys. Rev. Lett. \textbf{45}, 2081 (1980)}.

\bibitem{gpa}
R.~F.~C.~Vessot and M.~W.~Levine, \textit{Gravitational redshift space-probe experiment}, \href{https://ntrs.nasa.gov/search.jsp?R=19800011717}{NASA technical report NASA-CR-161409, 1979)}.

\bibitem{lisa-over} K.~Danzmann and the LISA study team, \emph{ LISA: laser interferometer space antenna for gravitational wave measurements},
\href{https://doi.org/10.1088/0264-9381/13/11A/033}{Class. Quantum Grav. \textbf{13}, A247 (1996)}.

 \bibitem{tdi} M.~Tinto and S.~V.~Dhurandhar, \emph{ Time-Delay Interferometry}, \href{https://link.springer.com/article/10.12942/lrr-2014-6}{Living Rev. Relat. \textbf{17}, 6 (2014)}.

\bibitem{qgo:03} G. M. Shore, \emph{Quantum gravitational optics}, \href{http://dx.doi.org/10.1080/00107510310001617106}{Contemp. Phys. \textbf{44}, 503 (2003)}.

\bibitem{ueno75prl}
Y. Ueno, \emph{On the wave theory of light in general relativity, I: path of light},
\href{https://doi.org/10.1143/PTP.10.442}{Progress of Theoretical Physics {\bf 10},
442 (1953)}.

\bibitem{harte:19} A. I. Harte, \emph{Gravitational lensing beyond geometric optics:
I. Formalism and observables}, \href{https://doi.org/10.1007/s10714-018-2494-x}{Gen. Relat. Gravit. \textbf{51}, 14 (2019)}.

\bibitem{qop} L. Mandel and E. Wolf, \textit{Optical Coherence and Quantum Optics},     (Cambridge University Press, Cambrdge, England, 1999).


\bibitem{will:93}
C.~M.~Will, \textit{Theory and Experiment in Gravitational Physics}, (Cambridge University Press, 1993).

\bibitem{wp:book} E. Poisson and C. W. Will, \textit{Gravity: Newtonian, Post-Newtonian, Relativistic}, (Cambridge University Press, 2014).

\bibitem{bggst:15}
A.~Brodutch, A.~Gilchrist, T.~Guff, A.~R.~H.~Smith, D.~R~Terno, \emph{ Post-Newtonian gravitational effects in optical interferometry}, \href{https://journals.aps.org/prd/abstract/10.1103/PhysRevD.91.064041}{Phys. Rev. D {\bfseries 91}, 064041 (2015)}.

\bibitem{gpa-proposal}
D.~Kleppner, R.~F.~C.~Vessot, N.~F.~Ramsey, \emph{An orbiting clock experiment to determine the gravitational red shift}, \href{https://doi.org/10.1007/BF00653616}{Astroph. Space Sci. \textbf{6}, 13 (1970)}.

\bibitem{Blanchet2001}
L.~Blanchet, C.~Salomon, P.~Teyssandier, and P.~Wolf,
\emph{Relativistic theory for time and frequency transfer to order $c^{-3}$}, \href{https://doi.org/10.1051/0004-6361:20010233}{A\&A 370, 320-329 (2001)}.

\bibitem{LaFitte2004}
C.~Le~Poncin-Lafitte, B.~Linet and P.~Teyssandier,
\emph{World function and time transfer: general post-Minkowskian expansions},
\href{https://doi:10.1088/0264-9381/21/18/012}{Class. Quantum Grav. {\bf 21} 4463-4483 (2004)}.

\bibitem{Hees2014}
A.~Hees, S.~Bertone, and C.~Le Poncin-Lafitte,
\emph{Relativistic formulation of coordinate light time, Doppler, and astrometric observables up to the second post-Minkowskian order},
\href{https://doi.org/10.1103/PhysRevD.89.064045}{Phys. Rev. D {\bf 89}, 064045 (2014)}.

\bibitem{cole75prl}
R.~Colella, A.~W. Overhauser, and S.~Werner, \emph{{Observation of gravitationally induced quantum interference}}, \href{https://doi.org/10.1103/PhysRevLett.34.1472}{Phys. Rev. Lett. \textbf{34}, 1472 (1975)}.

\bibitem{Rakhmanov2006} M.~Rakhmanov, \emph{Reflection of light from a moving mirror: derivation of the relativistic Doppler formula without Lorentz transformations}, \href{https://arxiv.org/abs/physics/0605100}{arXiv:physics/0605100 (2006)}.

\bibitem{zych11nco}
M.~Zych, F.~Costa, I.~Pikovski, and {\v{C}}.~Brukner, \emph{Quantum   interferometric visibility as a witness of general relativistic proper  time},
\href{https://doi.org/10.1038/ncomms1498}{Nature Commun. \textbf{2}, 505 (2011)}.


\bibitem{sat:12}
D.~Rideout, T.~Jennewein, G.~Amelino-Camelia, T.~F.~Demarie, B.~L.~Higgins, A.~Kempf, A.~Kent, R.~Laflamme,
 X.~Ma, R.~B.~Mann, E.~Mart{\'{\i}}n-Mart{\'{\i}}nez, N.~C. Menicucci, J.~Moffat, C.~Simon, R.~Sorkin, L.~Smolin and D.~R.~Terno,
  \emph{Fundamental quantum optics experiments conceivable with satellites-reaching relativistic distances and velocities},
   \href{http://iopscience.iop.org/article/10.1088/0264-9381/29/22/224011/meta}{Class. Quantum Grav. \textbf{29}, 224011 (2012)}.

\bibitem{us-paper}    D.~R.~Terno, F.~Vedovato, M.~Schiavon, A.~R.~H.~Smith, P.~Magnani, G.~Vallone, and P.~Villoresi,
\emph{Proposal for an optical test of the Einstein Equivalence Principle}, \href{https://arxiv.org/abs/1811.04835}  {arXiv: 1811.04835   (2018)}.

\bibitem{ll:2}
L. D. Landau and E. M. Lifshitz, \textit{The Classical Theory of
Fields}, (Butterworth-Heinemann, Amsterdam, 1980).

\bibitem{fl:82} F. Fayos and J. Llosa, \emph{Gravitational effects on the polarization plane},
\href{https://doi.org/10.1007/BF00756802}{Gen. Relativ. Gravit. {\bf 14}, 865 (1982)}.

\bibitem{bt:11} A. Brodutch and D. R. Terno, \emph{Polarization rotation, reference frames, and Mach’s principle},
\href{https://doi.org/10.1103/PhysRevD.84.121501}{Phys. Rev. D {\bf 84}, 121501(R) (2011)}.



 \bibitem{allen} G. Schubert and R. L. Walterscheid, \emph{Earth}, in A. N. Cox (ed.), \textit{Allen's  Astrophysical    Quantities}, 4th edition, (Springer, New York, 2002), p. 239.

\bibitem{gnss} N. Ashby, \textit{Relativity in GNSS}, in A. Ashtekar and V. Petkov (eds.), \emph{Springer Handbook of Spaceime} (Springer, Berlin, 2014), p.~509.





\bibitem{padova:16}
G.~Vallone, D.~Dequal, M.~Tomasin, F.~Vedovato, M.~Schiavon, V.~Luceri, G.~Bianco and P.~Villoresi, \emph{Interference at the single photon level along satellite-ground channels},
\href{https://doi.org/10.1103/PhysRevLett.116.253601}{Phys. Rev. Lett. {\bf 116}, 253601 (2016)}.


\end{thebibliography}
\end{document}